\documentclass[twocolumn]{aastex631}

\begin{document}


\title{Near and Mid UltraViolet Observations of X-6.3 flare on 22nd February 2024 recorded by the Solar Ultraviolet Imaging Telescope on board Aditya-L1}

\author[0000-0003-2215-7810]{Soumya Roy}
\affil{Inter-University Centre for Astronomy and Astrophysics, Post Bag 4, Ganeshkhind, Pune 411007}\email{soumyaroy@iucaa.in}
\author[0000-0003-1689-6254]{Durgesh Tripathi}
\affil{Inter-University Centre for Astronomy and Astrophysics, Post Bag 4, Ganeshkhind, Pune 411007}
\author[0000-0002-7276-4670]{Sreejith Padinhatteeri}
\affil{Manipal Centre for Natural Sciences, Manipal Academy of Higher Education, Karnataka, Manipal- 576104, India}
\author[0000-0001-5707-4965]{A. N. Ramaprakash}
\affil{Inter-University Centre for Astronomy and Astrophysics, Post Bag 4, Ganeshkhind, Pune 411007}
\author[0009-0002-0216-0545]{Abhilash R. Sarwade}
\affil{UR Rao Satellite Centre, Indian Space Research Organisation, Old Airport Road, Vimanapura PO, Bengaluru, 560017, India}
\affil{Indian Institute of Technology Guwahati, Guwahati, 781039, Assam, India}
\author[0000-0001-6866-6608]{V. N. Nived}
\affil{Inter-University Centre for Astronomy and Astrophysics, Post Bag 4, Ganeshkhind, Pune 411007}
\author[0000-0002-8560-318X]{Janmejoy Sarkar}
\affil{Inter-University Centre for Astronomy and Astrophysics, Post Bag 4, Ganeshkhind, Pune 411007}
\affil{Tezpur University, Napaam, Tezpur, Assam 784028}
\author[0000-0002-1282-3480]{Rahul Gopalakrishnan}
\affil{Inter-University Centre for Astronomy and Astrophysics, Post Bag 4, Ganeshkhind, Pune 411007}
\author[0009-0000-2781-9276]{Rushikesh Deogaonkar}
\affil{Inter-University Centre for Astronomy and Astrophysics, Post Bag 4, Ganeshkhind, Pune 411007}
\author[0000-0003-1406-4200]{K. Sankarasubramanian}
\affil{UR Rao Satellite Centre, Indian Space Research Organisation, Old Airport Road, Vimanapura PO, Bengaluru, 560017, India}
\affil{Center of Excellence in Space Sciences India, Indian Institute of Science Education and Research Kolkata, Mohanpur 741246, West Bengal, India}
\author[0000-0002-3418-8449]{Sami K. Solanki}
\affil{Max-Planck-Institut für Sonnensystemforschung, Justus-von-Liebig-Weg 3, 37077 Göttingen, Germany}
\author[0000-0001-5205-2302]{Dibyendu Nandy}
\affil{Center of Excellence in Space Sciences India, Indian Institute of Science Education and Research Kolkata, Mohanpur 741246, West Bengal, India}
\affil{Department of Physical Sciences, Indian Institute of Science Education and Research Kolkata, Mohanpur 741246, West Bengal, India}
\author[0000-0003-4653-6823]{Dipankar Banerjee}
\affil{Indian Institute of Space Science and Technology
Valiamala, Thiruvananthapuram - 695 547, Kerala, India}
\affil{Center of Excellence in Space Sciences India, Indian Institute of Science Education and Research Kolkata, Mohanpur 741246, West Bengal, India}

\begin{abstract}
Solar flares are regularly observed in extreme ultraviolet (EUV), soft X-rays (SXR), and hard X-rays (HXR). However, those in near and mid-UV are sparse. The Solar Ultraviolet Imaging Telescope (SUIT) onboard the Aditya-L1, launched on 2nd September, 2023 provides regular observations in the 200{--}400~{nm} wavelength range through eleven filters. Here, we report the observation of the X6.3 flare on Feb 22, 2024 using eight narrow band (NB) filters of SUIT. We have also used co-spatiotemporal observations from SDO/AIA, Solar Orbiter/STIX, GONG H$\alpha$, Aditya-L1/SoLEXS and GOES. We obtained light curves over the flaring region from AIA 1600, 1700~{\AA} and GONG H$\alpha$ and compared them with the disk-integrated lightcurve obtained from GOES and SoLEXS SXR and STIX HXR. We find that the flare peaks in SUIT NB01, NB03, NB04, and NB08 filters simultaneously with HXR, 1600, and 1700~{\AA} along with the peak temperature obtained from SoLEXS. In contrast, in NB02 and NB05, the flare peaks $\sim$~2 minutes later than the HXR peak, while in NB06 and NB07, the flare peaks $\sim$~3 minutes after the GOES soft X-ray peak. To the best of our knowledge, this is the first observation of a flare in these wavelengths (except in NB03, NB04 and NB05). Moreover, for the first time, we show the presence of a bright kernel in NB02. These results demonstrate the capabilities of SUIT observations in flare studies.
\end{abstract}

\section{Introduction} \label{sec:intro}

Solar flares are characterized by the explosive release of magnetic free energy and localized heating in the solar atmosphere. They manifest themselves differently across various wavelengths. While flare loops radiate mainly in soft X-rays (SXR) and the Extreme Ultraviolet (EUV) \citep[][]{tripathi04,tripathi06,fletcher11}, their footpoints are observed in hard X-rays (due to electron beams) and $\gamma$~rays (due to ion beams).  The corresponding white-light (WL) and near ultra-violet (NUV) counterparts have been demonstrated to be co-spatial and co-temporal with the sources of hard X-rays (HXR) \citep{hudson92, oliverso12, heinzel14}. The WL and NUV radiation is believed to arise due to two distinct mechanisms: (1) the Hydrogen recombination continua, namely Paschen and Balmer, produced in the chromosphere and (2) the photospheric continuum, mainly the $\mathrm{H^{-}}$ continuum, enhancement due to an increase in temperature below the temperature minimum \citep{ding03,ding07}, respectively. 

One of the major puzzling aspects regarding solar flares is the origin of the WL continuum. According to the standard model of flares, the energetic electrons are accelerated along the loops and are stopped via numerous collisions and thermalization when the ambient medium has enough density, usually known as the `thick target' model \citep[see e.g.,][and references therein]{brown73,benz17,fletcher11}. It is believed that such densities are already present within the chromosphere, explaining the co-spatiotemporal nature of WL and HXR emission. The theoretical estimations, on the other hand, predict that parts of the WL emission may originate in the upper photosphere \citep[][ and references therein]{ding07}, i.e., depths inaccessible to electrons in the standard `thick-target' model. 

There have been several attempts to modify the thick target model to account for various observations from solar flares, e.g., a warm target model \citep{kontar15,kontar19}, local re-acceleration of electrons \citep{brown73}, acceleration of deeper penetrating protons along with electrons. There were also attempts to explain the source of WL emission with photospheric `back-warming' \citep{metclaff90}, where the upper photosphere is ionized due to energy transported radiatively from the chromosphere. 

The above mentioned dichotomy suggests that further observations of flares in white-light, NUV and HXR is essential to come up to a firm conclusion. Among these, the least explored wavelength is the NUV, albeit some recent observations \citep[see e.g.,][]{heinzel14,kleint16,kleint17,kowalski17,kowalski19} from the Interface Region Imaging Spectrometer \citep[IRIS][]{iris}. 

In this paper, we discuss the first on-disk X-class flare recorded by the Solar Ultraviolet Imaging Telescope \citep[SUIT;][]{suit,suit_main,suit_test_calib} on board Aditya-L1 \citep{adityal1} on February 22, 2024. The rest of the paper is structured as follows. In \S\ref{obs}, we discuss the observations of the flare and the dataset used for its study. \S\ref{res} we discuss the data analysis and results. Finally, in \S\ref{sec:disc}, we summarize, conclude and discuss the importance of these observations in flare studies. 

\section{Observations}\label{obs} 
Active Region (AR) NOAA 13590 appeared on Feb. 18, 2024, at the northeast limb of the solar disk. On Feb. 22, 2024, it was located at N17E26 with $\beta\gamma\delta$ configuration and produced multiple flares, including an M~4.8, an X~1.7 and an X~6.3 flare. The X~6.3 flare SOL2024-02-22T22:08:00 is the first on-disk X-class flare that was observed by SUIT and is discussed here.

\begin{table}{
\centering
\begin{tabular}{lcccr}
\hline
Filter ID 	& Central 		    & FWHM 		& Remarks\\
		& Wavelength	    &			&         \\
		& (nm)			    & (nm)		&          \\
\hline
NB01 		& 214.0             & 11  	    &Continuum\\
NB02 		& 276.7 		    & 0.4  		&Continuum\\
NB03 		& 279.6 			& 0.4  		&\ion{Mg}{2}~k\\
NB04 		& 280.3 			& 0.4		&\ion{Mg}{2}~h\\
NB05 		& 283.2 			& 0.4		&Continuum\\
NB06 		& 300.0   			& 1.0		&Continuum\\
NB07 		& 388.0   			& 1.0		&CN Band\\ 
NB08 		& 396.85 			& 0.1 		&\ion{Ca}{2}~h\\
BB01 		& 200{--}242 		& 42.0		&Herzberg Continuum\\
BB02 		& 242{--}300 		& 58.0		&Hartley Band\\
BB03 		& 300{--}360 		& 40.0		&Huggins Band\\
\hline
\end{tabular}
\caption{List of science filters available on SUIT. Note that NB and BB stand for narrowband and broadband, respectively.} \label{sc_comb_fil}}
\end{table}

SUIT carries 11 filters (8 narrow bands and 3 broad bands) tuned to observe the solar photosphere and chromosphere at various wavelengths in the 200-400~{nm} wavelength range (see Table~\ref{sc_comb_fil}, for further details, please refer to \cite{sc_filt}). SUIT provides images at a pixel size of 0.7{\arcsec} and provides both full disk and Region of Interest (RoI) observations according to the observational sequence being used. . Four out of eight narrow band filters are tuned to observe the \ion{Mg}{2} blue wing continuum (NB02), \ion{Mg}{2}~k (NB03), \ion{Mg}{2}~h (NB05) and \ion{Mg}{2} red wing continuum (NB05). The NB06 and NB07 continuum filters are towards the blue and red sides of the Balmer jump (364.5~nm), respectively. 

The X~6.3 flare started at 22:08~UT and peaked at 22:34~UT. It was observed in all eight narrowband filters with an approximate temporal cadence of $\approx$1~minute. To the best of our knowledge, this is the first spatially resolved observation of a flare in these passbands centred at 214 (NB01), 276.7 (NB02), 300 (NB06) and 388 (NB07). Note that flare observations have been taken in these wavelengths using Sun-as-a-star observations, e.g., with Proba-2/LYRA integrated irradiance measurements in 190 {--} 222~nm \citep{dominique18} and spectral observations in 350 {--} 440~nm from the horizontal solar telescope HSFA2 of the Ond\v{r}ejov Observatory \citep{kotrvc16}.

The flare discussed here was also observed by Aditya-L1/SoLEXS\footnote{The SUIT and SoLEXS data are publicly available through \href{https://pradan.issdc.gov.in}{ISRO Science Data Archive}}  \citep{solexs}, GOES/XRS \citep{xrs}, SDO/AIA \citep{sdo,aia}, Solar Orbiter/STIX \citep{so,stix,stix1}, GONG H$\alpha$ \citep{gong}. In this study, we have used AIA~1600 \& 1700~{\AA}, GONG~H$\alpha$ images and full disk integrated GOES SXR (1{--}8~{\AA}~$\simeq$ 12.4 {--} 1.55 keV), SoLEXS SXR (2{--}20~keV~ $\simeq$ 6.2 {--} 0.62~{\AA}) and STIX HXR (25{--}50~keV) along with SUIT observations.

\begin{figure*}[ht!]
    \centering
    \includegraphics[trim = {2cm 2.5cm 2cm 1.8cm}, clip, width=0.9\textwidth]{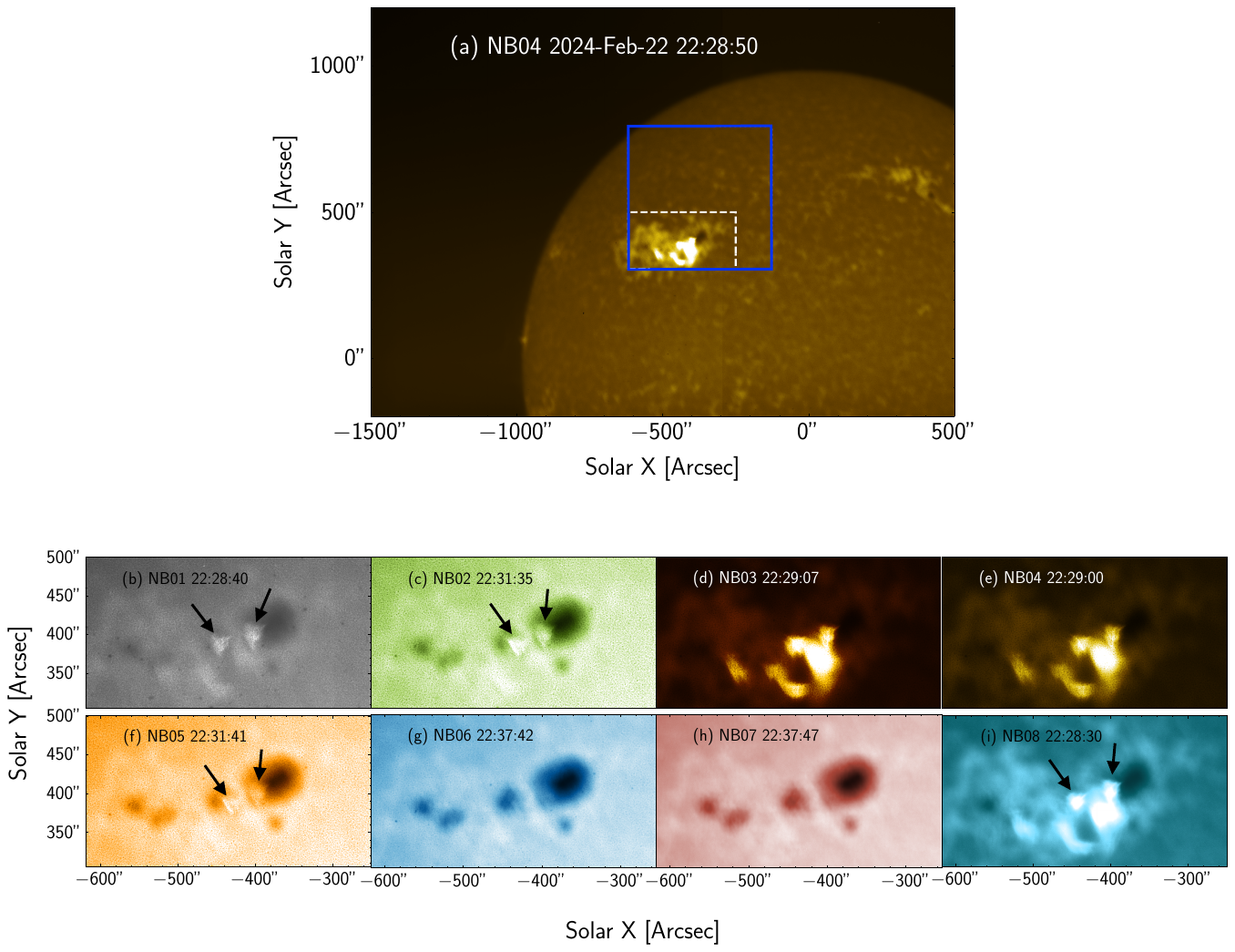} 
    \caption{(a) Parts of the full-disk (2k$\times$2k) observation in NB04 (\ion{Mg}{2}~h) showing the flaring region and sunspot. The over-plotted blue box shows the observed RoI, and the white dashed box locates the region used for the subsequent analysis. (b) SUIT Narrowband images recorded at the peak of flares in the respective filters. The size of the images corresponds to the white-boxed region shown in panel a. The arrows mark the two bright penumbral kernels.}
    \label{fig:flare_obs}
\end{figure*}
\section{Data Analysis and results}\label{res}

Figure~\ref{fig:flare_obs}.a displays a cutout of the full-disk (2k$\times$2k) image taken in the NB04 (\ion{Mg}{2}~h) filter. The over-plotted blue box shows the region of interest automatically defined by the onboard flare detection algorithm. As we can see, the auto-defined RoI fully covers the flare. The flare detection algorithm is designed to define the RoI, which is centred on the flaring region. In the current observation, we can see that the RoI is not properly centred at the flare. However, since then, various parameters of the flare detection and localization algorithm have been tweaked to achieve better detection and localization. The white dashed box locates the region we have considered for further study in this paper. Fig.~\ref{fig:flare_obs}.{b-i} displays the flare in each narrow band filter at their respective peak. The field of view (FOV) corresponds to the dashed white box in panel a. The flare is visible in all the narrow-band filters of SUIT, including the two flare kernels observed as penumbral brightenings, which are indicated by black arrows. 
\begin{figure*}
    \centering
    \includegraphics[trim={0.2cm 2.5cm 1cm 3.5cm},clip,width=\linewidth]{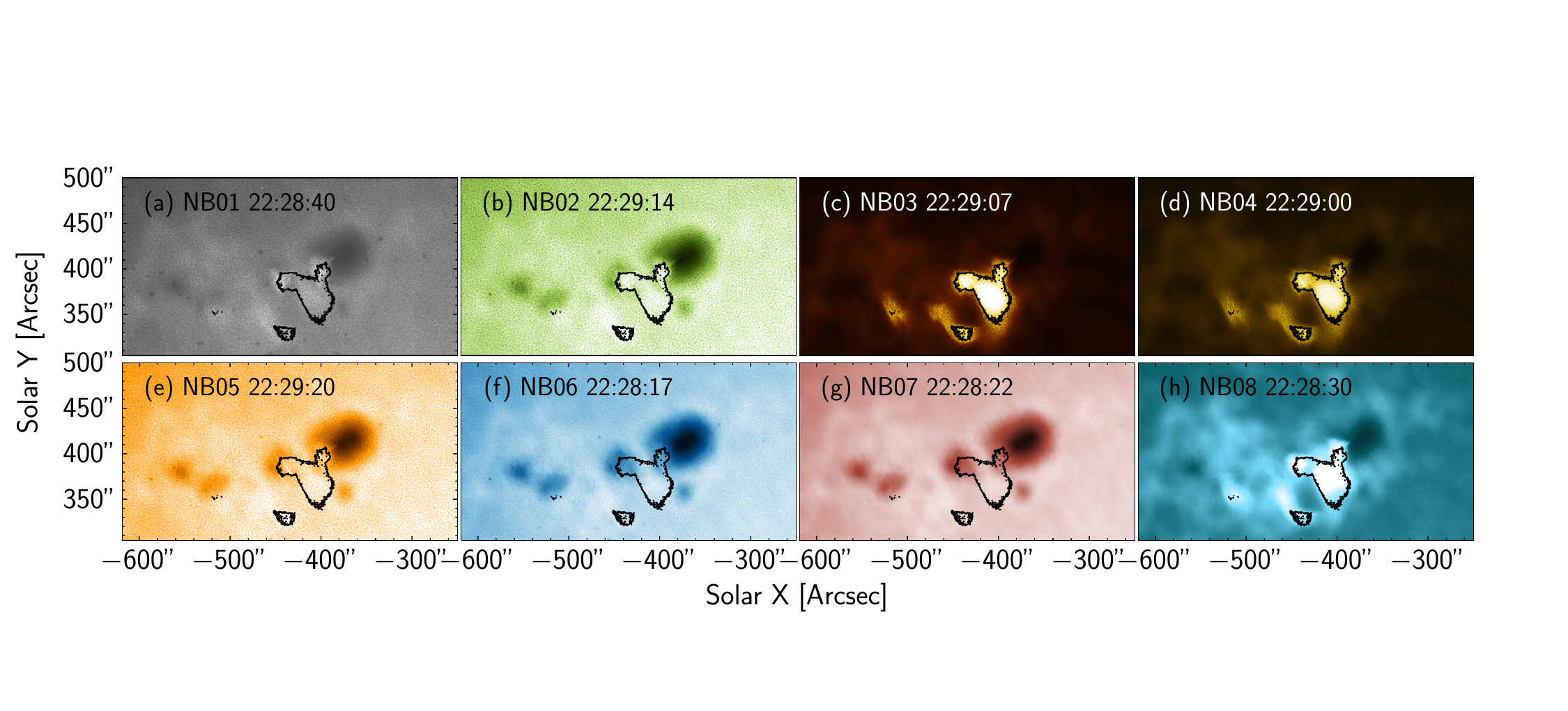}
    \caption{The flare as observed in all narrow band filters at the time of peak intensity in NB03. The over-plotted black contours show 60\% of the peak intensity observed in NB03.}
    \label{fig:flare_nb3_peak}
\end{figure*}
\begin{figure}
    \centering
    \hspace{-1cm}
   \includegraphics[trim={0.5cm 2.5cm 0.5cm 0cm},clip,width=\linewidth]{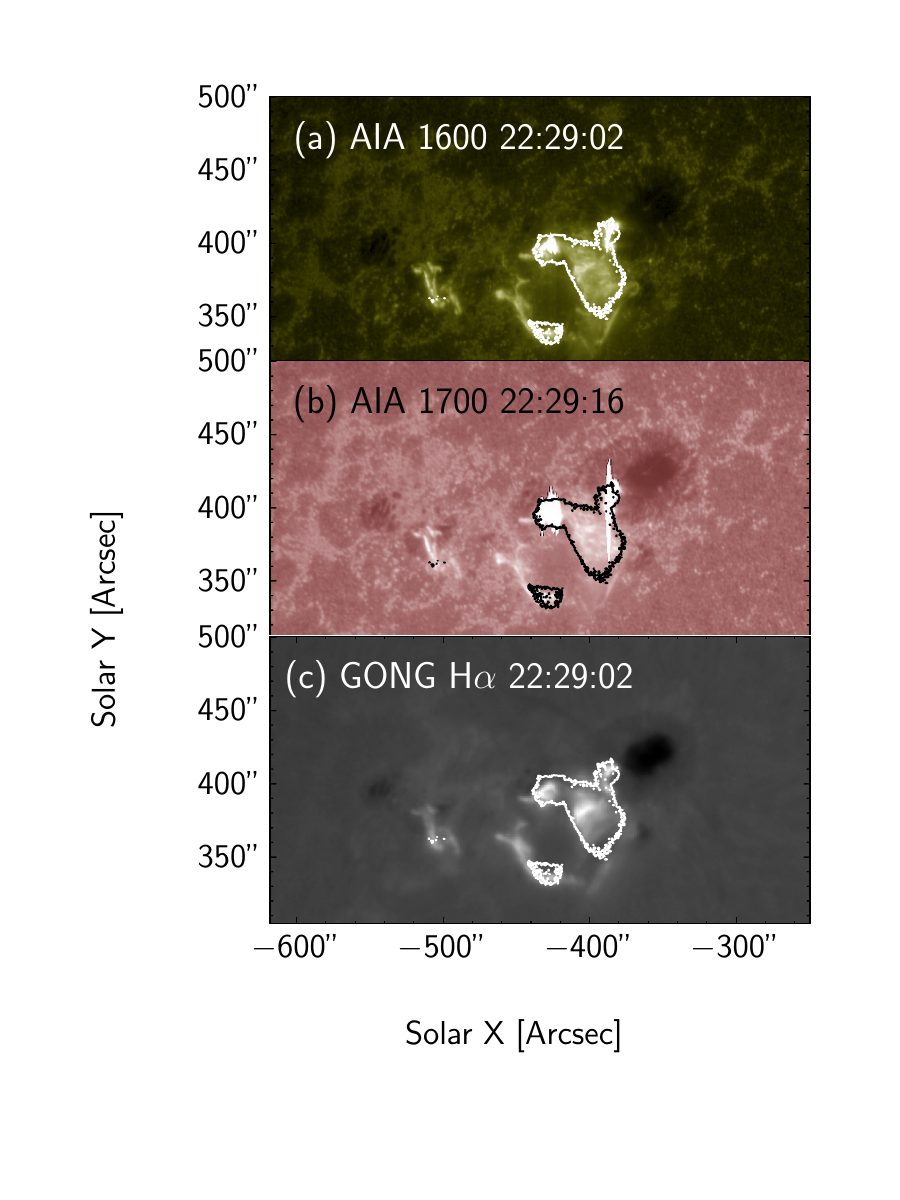}
    \caption{Co-aligned and co-registered AIA 1600, 1700~{\AA} and GONG H{$\alpha$} observation near the NB03 peak in a, b and c, respectively. The over-plotted contours show 60\% of the peak intensity observed in NB3.}\label{aia}
\end{figure}

In Fig.~\ref{fig:flare_nb3_peak} we plot the flare as observed in each narrow band filter at the peak intensity in NB03(\ion{Mg}{2}~k). The over-plotted black contours represent 60\% of the peak intensity observed in NB03. The features observed in NB08 (\ion{Ca}{2}~h) are very similar in morphology to those in NB03 and NB04. While there appears to be a hint of such features in other continuum channels, it is not clearly discernible. 

\begin{figure}[ht!]
    \centering
    \includegraphics[trim={0cm 0cm 0.5cm 0.5cm}, clip, width=\linewidth]{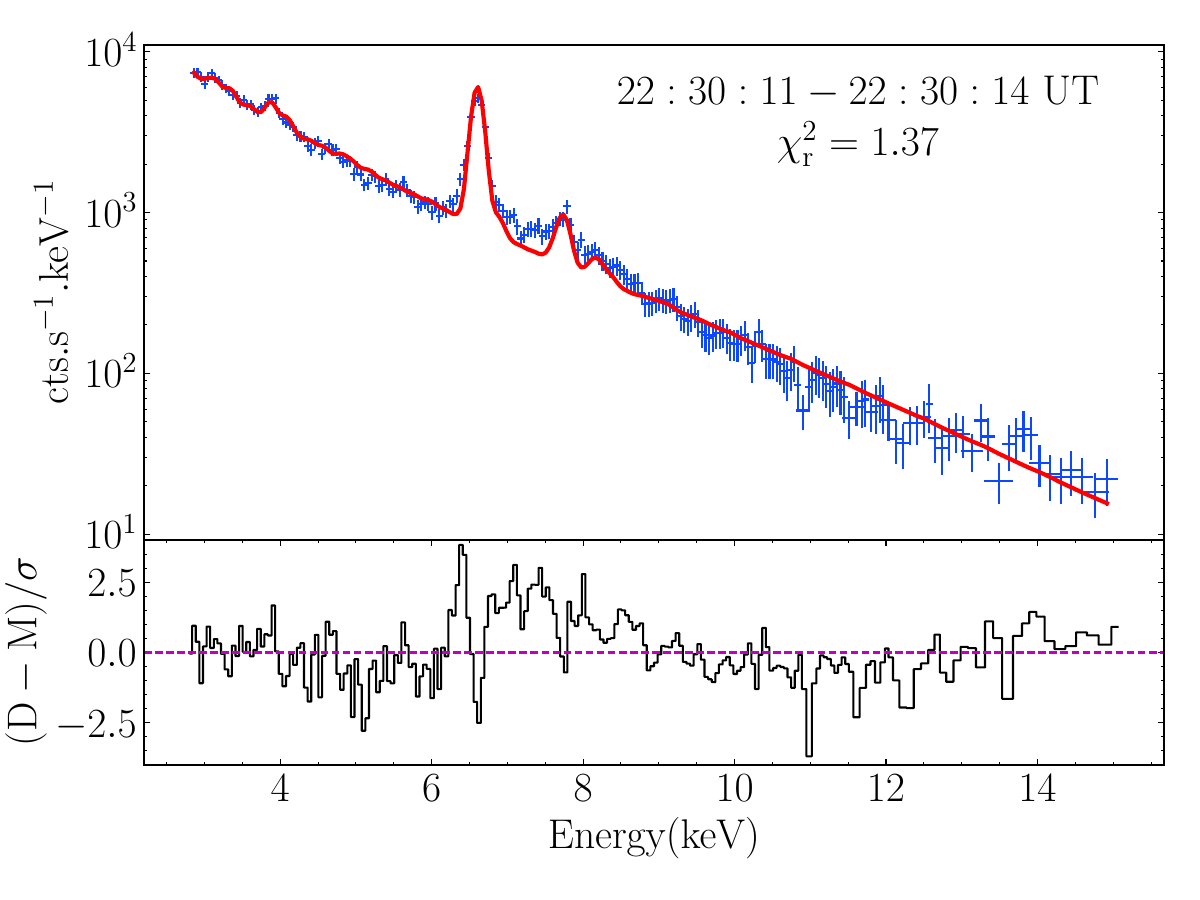}
    \caption{Reference fit (solid red) of the SoLEXS spectra (blue crosses) using an isothermal model. The bottom panel shows the fit residue.}
    \label{fig:sol_fit}
\end{figure}

\begin{figure*}
    \centering
    \includegraphics[width=\textwidth,trim={1cm 3cm 2cm 5.2cm},clip]{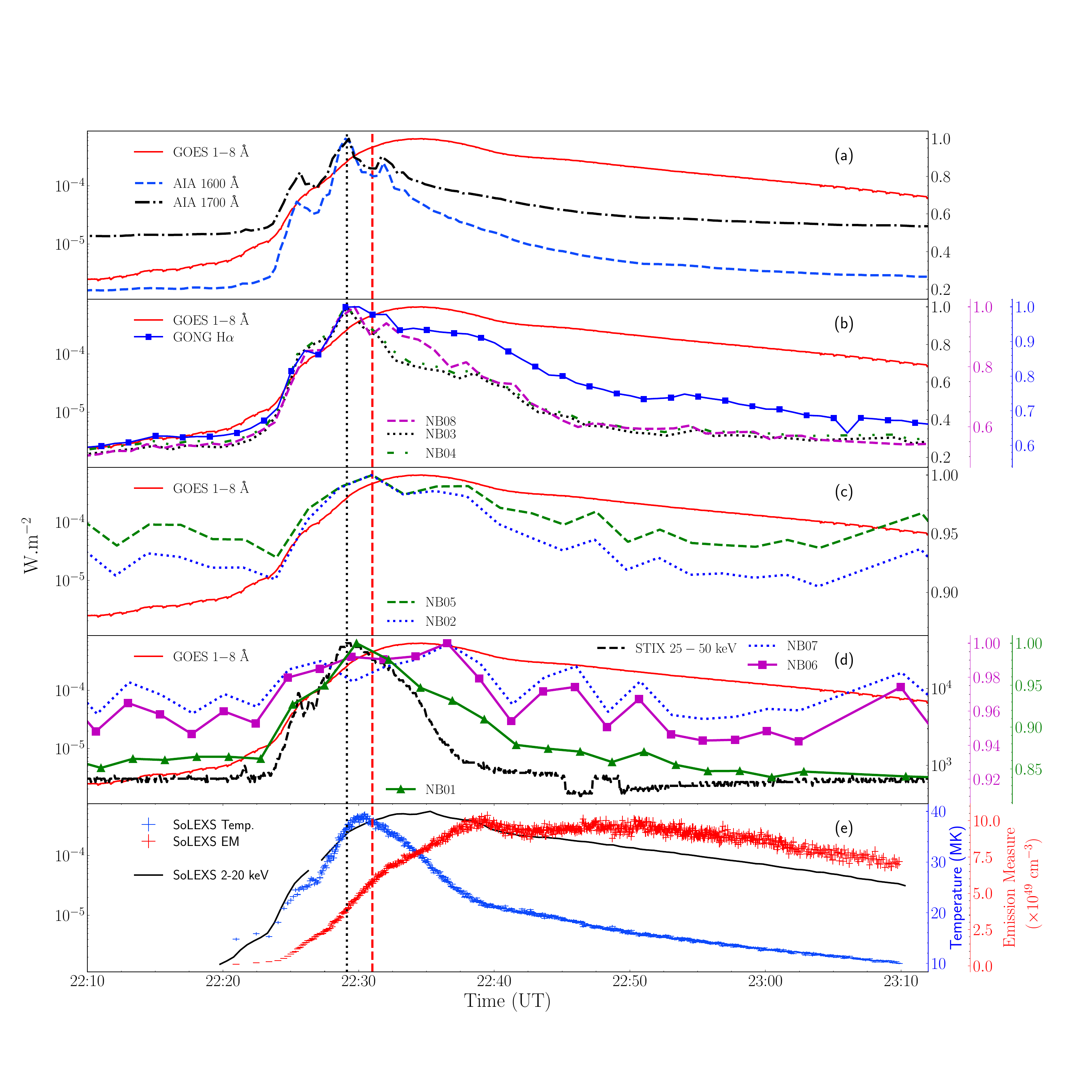}
    \caption{Light curves obtained over the contoured region from AIA 1600 (blue dashed), AIA 1700~{\AA} (black do-dashed), NB03 (black dotted), NB04 (green dot-dashed), NB08 (magenta dashed), NB02 (blue dotted), NB05 (green dashed), NB01 (green triangles), NB06 (magenta squares) and NB07 (blue dotted) as labelled. For comparison, we have also over-plotted the light curves obtained from coaligned GONG H-alpha observations (blue squares) in panel (b), full-sun integrated GOES/SXR in 1{--}8~{\AA} (red solid) and STIX/HRX (black dashed) in panel (d). In panel (f), we show the light curve obtained from SoLEXS 2{--}20 keV (black) observations and the calculated temperature (blue) and emission measure (red). The vertical dotted black line and dashed red line across the panels mark the peak time of NB03, NB04, NB08 and NB02, NB05, respectively.}
    \label{fig:flare_lc_suit}
\end{figure*}

To compare the SUIT observations with those from AIA and H$\alpha$, in Fig.~\ref{aia}, we plot AIA~1600 (panel a), AIA~1700 (panel b) and GONG-H$\alpha$ intensity recorded at nearly the same time as the intensity peak seen in NB03. The over-plotted contours are the same as those in Fig~\ref{fig:flare_nb3_peak}. These images clearly reveal signatures of the flare kernels in AIA UV and H$\alpha$ observations, co-spatial to the NB03 (\ion{Mg}{2}~k) observations.

In order to study the behaviour of the flare in SXR and obtain the temperature and emission measure as a function of time, we use temporally resolved spectra with an integration time of 3~s recorded by SoLEXS. We fitted the spectra using a forward-fitting procedure of \citep{sherpa_416, sherpa_paper} with an isothermal model, which is computed using the sunkit-spex \footnote{\url{https://github.com/sunpy/sunkit-spex}} python package. This package uses thermal bremsstrahlung for continuum and CHIANTI \citep{chianti, chianti10} atomic database for emission line calculation, assuming fixed coronal abundances. Note that when fitting spectra, the lower energy limit is set to 2.8~keV, while the upper energy limit is dynamically adjusted for each time-binned spectrum to ensure sufficient counts in the final detector channel. Moreover, the high-energy channels are grouped to maintain a good signal-to-noise ratio in each spectral channel. In addition, the isothermal model was modified locally to allow abundances as free parameters. Fig.~\ref{fig:sol_fit}, we show a reference fit to the spectrum obtained over a time interval of 22:30:11 to 22:30:14. We find that the fitted peak temperature is $\sim~(39.1~\pm~0.6)$ MK and the emission measure peak is $\sim~(10.3~\pm~0.2)~\times~10^{49}~\mathrm{cm^{-3}}$.

To study the temporal evolution of the flare, we plot the lightcurves in Fig.~\ref{fig:flare_lc_suit} obtained from various filters of SUIT, GOES, SoLEXS, STIX, AIA 1600 \& 1700, and GONG H$\alpha$, as labelled. Note that the GOES and STIX lightcurves are full Sun integrated, whereas those for other observations are obtained over the contoured region in Figs.~\ref{fig:flare_nb3_peak}~\&~\ref{aia}, and normalized to their respective max intensities. The vertical dotted black line across all the panels locates the peak intensity in NB03. 

The light curves reveal that the flare peaks in all the SUIT channels (except NB06 and NB07), AIA 1600 \& 1700~{\AA}, H$\alpha$ and STIX before it achieves its maximum intensity in GOES \& SoLEXS. The STIX peak coincides with that in SUIT NB01, NB03, NB04, NB08 as well as AIA and H$\alpha$. Note that SUIT NB03, NB04 and NB08 primarily observe the chromospheric signatures, similar to GONG H$\alpha$. Fig.~\ref{fig:flare_lc_suit}.b, shows that the lightcurves of NB03, NB04, and NB08 peak at the same time around $\sim$ 22:29~UT, which is similar to H$\alpha$. However, NB08 and {\it GONG}-H$\alpha$ show less contrast variation ($\sim$ 40\% change for NB08 (\ion{Ca}{2} h) and GONG H$\alpha$, compared to the $\sim$ 80\% change for NB03 (\ion{Mg}{2} k) and NB04 (\ion{Mg}{2} h)). This may be attributed to the larger enhancement in the region around the AR in NB08 (\ion{Ca}{2}) and H$\alpha$ than that in \ion{Mg}{2}.

Figure~\ref{fig:flare_lc_suit}.c shows that the lightcurves of SUIT NB02 and NB05 (blue and red wing of \ion{Mg}{2}) peak about 2 minutes after the NB03 peak (and the STIX hard X-ray peak). The black vertical dotted line and red vertical dashed line across all the panels locate the peak intensity in NB03 and NB05, respectively. This is also apparent in the images shown in Fig~\ref{fig:flare_obs}. Fig.~\ref{fig:flare_lc_suit}.d shows that the STIX Hard X-ray light curve (25{--}50 keV) peaks around the same time as NB03, NB04, NB08, NB01 and GONG-H$\alpha$. Moreover, unlike other lightcurves, the NB06 and NB07 lightcurves do not show a gradual increase and decrease (see Fig.~\ref{fig:flare_lc_suit}.d). Instead, they exhibit a rather sharp rise, albeit very small ($\sim$ 0.03\%), in intensity during the impulsive phase of the flare. Fig.~\ref{fig:flare_lc_suit}.e displays the 2{--}20~keV SoLEXS lightcurve (black), temperature (blue) and emission measure (red). Note the remarkable similarities between the SoLEXS lightcurve and GOES 1{--}8~{\AA} lightcurve. The temperature peak of the flare occurs at the same time as the Hard X-ray peak. The emission measure, however, peaks after the temperature peak ($\sim$~9 minutes), as expected \citep[see e.g.,][]{raftery09,sadykov19}.

\section{Summary and Discussion}\label{sec:disc}

In this letter, we have reported a multi-wavelength study of the X-6.3 flare observed on Feb 22, 2024. For this purpose, we have used recorded by SUIT, SolEXS, AIA, STIX and GONG. Note that this is the first on-disk X-class flare observed by SUIT. Below we summarize the obtained results.

\begin{itemize}
    \item The flare is observed in all narrow-band filters of SUIT. The flare was also observed in AIA 1700 and 1600, GONG~H$\alpha$, in SXR with SoLEXS \& GOES and STIX HXR.
    \item The flare peaked in all SUIT channels (except NB06 and NB07) as well as in AIA 1600 \& 1700~{\AA}, GONG~H$\alpha$ and STIX HXR $\sim$ 5 minutes prior to the GOES and SolEXS SXR peak. In NB06 and NB07, the flare peaks were observed $\sim$ 2 minutes after the GOES and SoLEXS SXR peak.
    \item The flare peak in NB01, NB03, NB04 and NB08 coincided with that in AIA 1600 \& 1700~{\AA} and STIX 25{--}50~keV observations. Moreover, the temperature obtained from SoLEXS peaked at the same time.
    \item The flare attained its peak in NB02 and NB05, i.e., the blue and red wings of the \ion{Mg}{2} lines, about 2 minutes after it did in NB03. 
\end{itemize}

The results demonstrate that the flare signatures detected by SUIT in all its narrowband channels are very tightly related to HXR observations recorded by STIX. The observations of flare kernels and enhancement in the intensity in the red wing of the \ion{Mg}{2} window, i.e., NB05 is similar to those observed by \cite{heinzel14,kleint16,kleint17,kowalski19,kowalski17} using the observations recorded IRIS. For the flare studied here, we also detect the bright kernels and intensity enhancements in the blue wing continuum of the Mg window, i.e., in the NB02 filter. To the best of our knowledge, this is the first such observation in the blue wing continuum of \ion{Mg}{2}. It is worth mentioning that enhancements of the Balmer continuum significantly closer to the \ion{Mg}{2} k line ($\sim$~2782.56 {--} 2795~{\AA}) have been reported before by \cite{joshi21}. Our observations are further on the bluer side $\sim$ 2767~{\AA}. The simultaneous appearance of the NUV intensity peak and soft X-ray temperature peak suggests that the observed UV emission is the heated-up plasma just at the stage of particle deceleration and plasma heating \citep{fletcher11, hudson06}.

It is important to highlight that in the observations reported in previous studies \citep[e.g.,][]{heinzel14,kleint17,kowalski19}, the appearance of flare kernel and intensity enhancement in IRIS SJI recorded at 2832~{\AA} (similar to SUIT NB05 filter) corresponded to a plethora of photospheric absorption lines, mostly \ion{Fe}{2}, turning into emission, which was attributed to photospheric heating. Note that for weaker flares, these photospheric lines may not go into emission. However, the presence of localized brightening has also been demonstrated to be due to continuum enhancement \citep{joshi21}. Since our flare is a strong X6 flare, the brightening observed by SUIT in NB05 is most likely from similar photospheric lines. Unfortunately, for the flare studied here, IRIS observed the eastern edge of the flaring region and did not observe the flare. Therefore, no such conformation was possible. 

The close correlation between the lightcurves of NB06 and NB07 is further intriguing as the formation of these continuum are from very different sources. NB06 (300 nm) possibly shows the Balmer continuum enhancement as it is on the blue side of the Balmer jump at $\sim$ 364.6 nm, while the NB07 (388 nm) is possibly exhibiting the enhancements in the photosphere through back warming of the enhanced Balmer continuum. Such a scenario may explain the observed correlation between the NB06 and NB07 light curves. Since this the first time ever a flare observation is being reported in these channels and further studies including numerical simulations are required to obtain a firm conclusion.

The results obtained in this letter demonstrate the capabilities of SUIT in flare observations and will help us understand the emission observed in NUV. Such observations will lead to constrain the physics of flares and derive spectral energy distribution in solar flares.
\begin{acknowledgments}
We thank the referee for several insightful  comments and suggestions that helped us improve our manuscript.  We thank Lyndsay Fletcher for her valuable suggestions and discussions. SUIT is built by a consortium led by the IUCAA, Pune, and supported by ISRO as part of the Aditya-L1 mission. The consortium consists of CESSI-IISER Kolkata (MoE), IIA, MAHE, MPS, USO/PRL, and Tezpur University. Aditya-L1 is an observatory class mission which is funded and operated by the ISRO. The mission was conceived and realized with the help from all ISRO Centres and payloads were realized by the payload PI Institutes in close collaboration with ISRO and many other national institutes - IIA; IUCAA; LEOS of ISRO; PRL; URSC of ISRO; VSSC of ISRO. SoLEXS is designed and developed at the Space Astronomy Group of URSC, ISRO with the help from various entities within URSC. We acknowledge the use of data from the Aditya-L1, the first solar mission of the ISRO, archived at the ISSDC. CHIANTI is a collaborative project involving George Mason University, the University of Michigan (USA), University of Cambridge (UK) and NASA Goddard Space Flight Center (USA). SP acknowledges the Manipal Centre for Natural Sciences, Centre of Excellence, and Manipal Academy of Higher Education (MAHE) for facilities and support. SKS acknowledges the funding from the European Research Council (ERC) under the European Union's Horizon 2020 research and innovation programme (grant agreement No. 101097844 — project WINSUN). The AIA data used here are courtesy of SDO (NASA) and the AIA and HMI consortium. This work utilizes GONG data obtained by the NSO Integrated Synoptic Program, managed by the NSO, which is operated by AURA, Inc. under a cooperative agreement with the NSF and with contribution from the NOAA. The GONG network of instruments is hosted by the BBSO, HAO, Learmonth Solar Observatory, USP, IAS Canarias, and Cerro Tololo Interamerican Observatory. The GOES 8-16 X-ray/ magnetic field/ particle data are produced in real-time by the NOAA Space Weather Prediction Center (SWPC) and are distributed by the NOAA NGDC. Solar Orbiter is a space mission of international collaboration between ESA and NASA, operated by ESA. The STIX instrument is an international collaboration between Switzerland, Poland, France, Czech Republic, Germany, Austria, Ireland, and Italy. The STIX data used in this research is and necessary analysis tools are publicly available from STIX data center \citep{stix_d_cen}.This research used version 6.0.5 \citep{sunpy_ver} of the SunPy open-source software package \citep{sunpy20} and PYTHON packages NumPy \citep{numpy}, Matplotlib \citep{matpltolib}, SciPy \citep{scipy}.
\end{acknowledgments}
\facilities{SUIT/Aditya-L1, SoLEXS/Aditya-L1, SDO, SolO, GOES, GONG}
\bibliography{main}{}
\bibliographystyle{aasjournal}
\end{document}